# Ni-assisted transformation of graphene flakes to fullerenes

*Irina V. Lebedeva[1,*], Andrey A. Knizhnik[1,2,*], Andrey M. Popov[3,*], and Boris V. Potapkin[1,2]*

[1]Kintech Lab Ltd., Kurchatov Square 1, Moscow, 123182, Russia,

[2]National Research Centre "Kurchatov Institute", Kurchatov Square 1, Moscow, 123182, Russia,

[3]Institute of Spectroscopy of Russian Academy of Sciences, Fizicheskaya Street 5, Troitsk, Moscow Region, 142190, Russia

*Address correspondence to lebedeva@kintechlab.com, am-popov@isan.troitsk.ru, knizhnik@kintechlab.com

**ABSTRACT** Transformation of graphene flakes to fullerenes assisted by Ni clusters is investigated using molecular dynamics simulations. The bond-order potential for Ni-C systems is developed. The potential reproduces the experimental and first-principles data on the physical properties of pure Ni as well as on relative energies of carbon species on Ni surfaces and in Ni bulk. The potential is applied for molecular dynamics simulations of the transformation of graphene flakes consisting of 50 – 400 atoms with and without Ni clusters attached. Free fullerenes, fullerenes with Ni clusters attached from outside and fullerenes encapsulating Ni clusters (Ni endofullerenes) are observed to form in the presence of Ni clusters consisting of 5–80 atoms. Moreover, a new type of heterofullerenes with a patch made of a Ni cluster is found to form as an intermediate structure during the transformation. The Ni clusters are shown to reduce the activation energy for the graphene-fullerene transformation from 4.0 eV to 1.5 – 1.9 eV, providing the decrease of the minimal temperature at which such a transformation can be observed experimentally from about 1400 K for free graphene flakes to about 700 – 800 K. While the transformation of free graphene flakes is found to occur through formation of chains of two-coordinated carbon atoms at the flake edges, the mechanism of the Ni-assisted graphene-fullerene transformation is revealed to be based on the transfer of carbon atoms from the graphene flake to the Ni cluster and back. The way of controlled synthesis of endofullerenes with a transition metal cluster inside and heterofullerenes with a transition metal patch is also proposed.

## 1. INTRODUCTION

Carbon nanostructures (fullerenes, carbon nanotubes and graphene) which are considered as a basis for future nanoelectronics can exhibit different properties depending on their atomic structure. To produce nanostructures appropriate for a particular type of applications, the ways to control the structure of these nanoobjects should be elaborated. Progress in this field is impossible without understanding the mechanisms and kinetics of structural transformations of carbon nanostructures.

Though carbon nanostructures are successfully synthesized experimentally, the mechanisms of their formation are still unclear. A present-day understanding of the mechanisms of formation of fullerenes in chaotic carbon plasma is discussed in a review.[1] New ways to produce fullerenes which allow controlling their structure during the synthesis were realized recently.[2,3] The cyclodehydrogenation of polyaromatic precursors on the platinum surface was used to produce the $C_{60}$ fullerene and the $C_{57}N_3$ triazafullerene.[2] The transformation of a small graphene flake to a fullerene under the action of an electron beam in the transmission electron microscope was demonstrated.[3] The following mechanism of the transformation was proposed.[3] First, etching of the graphene flake by the electron beam leads to formation of a notch in the flake. Then the flake relaxes with zipping the notch edges and curving into a bowl-shaped structure. The sequence of formation of notches followed by the structure relaxation finally leads to folding of the graphene flake into the fullerene. It should be emphasized that etching of graphene flakes by an electron beam is crucial for the transformation to fullerenes at low temperatures. It should also be noted that fullerenes obtained in this way can contain structural defects such as polygons other than hexagons and pentagons. Below we use the term "fullerene" not only in the narrow sense for perfect fullerenes, i.e. carbon cages consisting of three-coordinated carbon atoms arranged in hexagons and pentagons, but also for fullerenes incorporating structural defects.

The thermally-activated graphene-fullerene transformation was previously studied using molecular dynamics (MD) simulations.[4,5] The MD simulations revealed that at temperatures 3000 – 3500 K folding of graphene flakes consisting of 100 – 700 atoms occurs within several nanoseconds. The transformation was found to proceed through the formation of numerous polygons different from hexagons at the edges of the bond network of the flake and the detailed kinetics of these polygons was investigated. In the present paper we propose that metal clusters can be used as catalysts providing a substantial decrease of the temperature of the graphene-fullerene transformation. We also suggest that the graphene-fullerene transformation assisted by metal clusters can make it possible to synthesize new endofullerenes and heterofullerenes which are of interest for application in nanoelectronic devices and magnetic data storage.[6]

A wide set of experiments show that transition metals have an extremely high catalytic activity for C-C bond reorganization and formation of various carbon nanostructures. Among such experiments, there are numerous studies on synthesis of carbon nanotubes[7–10] and graphene,[11] transformation of single-walled nanotubes encapsulating fullerenes with Os complexes attached to their exterior[12] or ferrocenes[13] into double-walled nanotubes, nanoprotrusion formation on a single-walled nanotube encapsulating fullerenes with Re complexes attached to their exterior.[14] Nickel is one of the most active catalysts for growth[7–11] and controlled cutting[15] of carbon nanostructures. The distance between nearest atoms in Ni (2.49 Å) is close to the lattice constant of graphene (2.46 Å), making the growth of graphene and related structures on Ni (111) surfaces the easiest. Therefore we start our consideration of the catalytic effect of metal clusters on the graphene-fullerene transformation from Ni clusters. We perform MD simulations to investigate the process of the Ni-assisted graphene-fullerene transformation and to study possible Ni-C nanostructures that can be obtained in the result of this process.

Predictive atomistic modeling relies on the use of accurate interatomic potentials. While embedded-atom (EAM) potentials[16–18] are optimal for simulations of metals, reactive bond-order potentials[19–22] are widely used for simulations of covalently bonded systems. Unfortunately, a unified description of mixed carbon-metal systems with the same simple empirical function is not straightforward. Recently a modified embedded atom (MEAM) potential for Ni-C systems was suggested with the parameters of the Ni-C interaction fitted to the properties of Ni carbide.[23] Bond-order potentials for metal-carbon systems were developed[24–26] on the basis of first-principles calculations for small clusters ($MC_n$ and $M_n$ (n = 1 – 3) clusters,[24] M= La, Sc, Ni; $MC_n$ and $M_n$ (n = 1 – 4) clusters,[25] M = Ni, Co, Fe; $NiC_2H_4$ and $NiC_{16}H_{10}$ clusters[26]). Based on the experimental data for Pt and the results of first-principles calculations for different Pt carbides, the bond-order Brenner potential[19] was extended to Pt-C systems.[27] The bond-order potential for Fe-C systems[28] was fitted to the experimental and first-principles data for Fe carbides. However, all these potentials[23–28] have been poorly tested with regard to relative energies of carbon species on the metal surface and in the bulk, which is crucial for the adequate description of catalyzed growth and structural transformations of carbon nanostructures. In addition to the above semi-empirical potentials specific for particular systems, the description of metal-carbon interactions has been added to universal force fields (*e.g.*, ReaxFF[29,30]). A tight-binding potential for carbon interacting with transition metals[31,32] has been also developed. Nevertheless, such many-parameter force fields or tight-binding potentials are complicated and computationally expensive. The use of more efficient semi-empirical potentials is preferred for large-scale and/or long-time simulations.

In the present paper we extend the Brenner potential[19] to describe Ni-C systems. The parameters of the potential for the Ni-Ni interaction are fitted to experimental data on the lattice constant, cohesive energy and elastic properties of fcc Ni and to the characteristics of the nickel dimer obtained by density functional theory (DFT) calculations. The parameters for the Ni-C interaction are fitted to the relative energies of carbon adatoms at different sites of the Ni (111) surface, of atoms in $C_6$ rings and in graphene on the Ni (111) surface and of carbon interstitials in the subsurface layer and in the bulk obtained by the DFT calculations. Thus, the potential is specifically developed for simulations of the catalytic effect of Ni surfaces and clusters on the formation and transformations of carbon nanostructures.

The paper is organized as follows. In Sec. 2 we describe the developed potential and test it for Ni and Ni–C systems. In Sec. 3 we present the results of MD simulations of the transformation of graphene flakes to fullerenes at high temperatures. First, the results for free graphene flakes are given. Then the influence of Ni clusters is discussed. Our conclusions are summarized in Section 4.

## 2. DEVELOPMENT OF POTENTIAL FOR NI-C SYSTEMS

**2.1. Empirical Potential.** We have extended the Brenner potential[19] to describe Ni-Ni and Ni-C interactions. The same as in the original Brenner potential[19], the energy of the system is represented as

$$E_b = \sum_i \sum_{j(>i)} E_{ij}, \qquad (1)$$

where the energy $E_{ij}$ of the bond between atoms $i$ and $j$ separated by the distance $r_{ij}$ is given by the sum of repulsive and attractive terms

$$E_{ij} = V_R(r_{ij}) - \bar{b}_{ij} V_A(r_{ij}). \qquad (2)$$

The repulsive interaction is determined by the two-body function

$$V_R(r_{ij}) = f_{ij}(r_{ij}) A_{ij} \exp(-\lambda_{1,ij} r_{ij}), \qquad (3)$$

where the cutoff function $f_{ij}(r)$ has the form

$$f_{ij}(r) = \begin{cases} 1, & r < R_{ij}^{(1)} \\ \dfrac{1}{2}\left[1 + \cos\left[\dfrac{\pi(r - R_{ij}^{(1)})}{(R_{ij}^{(2)} - R_{ij}^{(1)})}\right]\right], & R_{ij}^{(1)} \leq r \leq R_{ij}^{(2)} \\ 0, & r > R_{ij}^{(2)} \end{cases} \qquad (4)$$

The attractive interaction is described by the two-body function

$$V_A(r_{ij}) = f_{ij}(r_{ij}) B_{ij} \exp(-\lambda_{2,ij} r_{ij}) \qquad (5)$$

multiplied by the function $\bar{b}_{ij}$ which describes the dependence of the interaction energy on the local coordination. The empirical bond order function $\bar{b}_{ij}$ is given by the sum of the average of the terms $b_{ij}$ and $b_{ji}$ corresponding to each atom in the bond and of the additional correction function $F_{ij}$, which is used to account for conjugated versus non-conjugated bonding and to avoid the overbinding of radicals,

$$\bar{b}_{ij} = (b_{ij} + b_{ji})/2 + F_{ij}\left(N_{ij}^{C}, N_{ji}^{C}, N_{ij}^{conj}\right), \tag{6}$$

where $N_{ij}^{C}$ is the number of carbon atoms bonded to atom $i$ except for atom $j$ and $N_{ij}^{conj}$ is used to determine whether the bond between atoms $i$ and $j$ is a part of a conjugated system. The function $F_{ij}$ is non-zero only for bonds between two carbon atoms and takes the same values as in the original Brenner potential[19].

The bond order function $b_{ij}$ for each atom in the bond is determined by

$$b_{ij} = \left[1 + \sum_{k(\neq i,j)} G_{ijk}(\theta_{ijk}) f_{ik}(r_{ik}) \exp\left[\alpha_{ijk}\left((r_{ij} - R_{ij}^{(e)}) - (r_{ik} - R_{ik}^{(e)})\right)\right]\right]^{-\delta}, \tag{7}$$

where $R_{ij}^{(e)}$ is the equilibrium distance between atoms $i$ and $j$, $\theta_{ijk}$ is the angle between the bonds between atoms $i$ and $j$ and atoms $i$ and $k$ and $\delta$ is taken equal to 0.5 for all atoms. The function $G_{ijk}(\theta)$ is taken in the form

$$G_{ijk}(\theta) = a_{ijk}\left[1 + \frac{c_{ijk}^2}{d_{ijk}^2} - \frac{c_{ijk}^2}{d_{ijk}^2 + (1 + \cos\theta)^2}\right]. \tag{8}$$

As opposed to the original Brenner potential[19], we assume that the parameters of the function $G_{ijk}(\theta)$ $a_{ijk}$, $c_{ijk}$ and $d_{ijk}$ depend on types of all three atoms $i$, $j$ and $k$.

The numbers $N_{ij}^{C}$ and $N_{ij}^{conj}$ are found as

$$N_{ij}^{C} = \sum_{C\,k(\neq j)} f_{ik}(r_{ik}), \tag{9}$$

$$N_{ij}^{conj} = 1 + \sum_{C\,k(\neq i,j)} f_{ik}(r_{ik}) F_0(N_{ki}^{C}) + \sum_{C\,l(\neq i,j)} f_{jl}(r_{jl}) F_0(N_{lj}^{C}), \tag{10}$$

where

$$F_0(x) = \begin{cases} 1, & x \leq 2 \\ [1 + cos(\pi(x-2))]/2, & 2 < x < 3 \\ 0, & x \geq 3 \end{cases} \quad (11)$$

The parameters for the C-C interaction are taken the same as in the second set of the parameters of the original Brenner potential[19] (Table 1 – Table 3), while the parameters for Ni-Ni and Ni-C interactions are fitted to the experimental data and the results of first-principles calculations.

**2.2. DFT calculations.** To fit the parameters of the potential for Ni and Ni-C systems and to test the potential we have performed DFT calculations for carbon species on Ni (111) and (113) surfaces (Figure 1), bulk Ni phases and Ni dimer. The VASP code[33] with the Perdew-Wang exchange-correlation functional[34] is used. The basis set consists of plane waves with the maximum kinetic energy of 358 eV. The interaction of valence electrons with atomic cores is described using ultrasoft nonlocal pseudopotentials.[35] In calculations for the Ni (111) and (113) surfaces, the periodic boundary conditions are applied to a 3x3x4 model cell. The metal slabs are separated by a 10 Å vacuum gap. Integration over the Brillouin zone is performed using the Monkhorst-Pack method[36] with a 7x7x1 k-points sampling. The calculations of the properties of bulk Ni phases are performed for 1x1x1 model cells. 13x13x13 and 17x17x17 k-point grids are used for fcc and bcc Ni, respectively. In the calculations for carbon interstitials in fcc Ni, a 3x3x3 model cell and a 3x3x3 k-point grid are considered. The calculations for Ni dimer are performed using a 15 Å x 15 Å x 15 Å model cell and a single G-point. The structures are optimized until the maximal residual force acting on each atom is less than 0.03 eV/Å.

The results of our DFT calculations are summarized in Table 4 and Table 5. It is seen from Table 4 that the calculated lattice constant $a_0$, cohesive energy $E_{coh}$ and bulk modulus $B$ of fcc Ni as well as the calculated surface energy $E_{(111)}$ of the (111) surface and the relative cohesive energy $\Delta E_{coh}$ of bcc Ni are in reasonable agreement with the experimental data.[17,37–41] The calculated relative energies of carbon atoms on the Ni surfaces and in the Ni bulk are consistent with previous DFT calculations[42–48] (Table 5).

**Table 1.** Two-body parameters of the potential.

| Parameters | C-C | C-Ni | Ni-Ni |
|---|---|---|---|
| $A$ (eV) | 2606 | 1866 | 1473 |
| $B$ (eV) | 1397 | 184.6 | 61.24 |
| $\lambda_1$ (Å$^{-1}$) | 3.2803 | 3.6768 | 3.2397 |
| $\lambda_2$ (Å$^{-1}$) | 2.6888 | 1.8384 | 1.2608 |
| $R^{(1)}$ (Å) | 1.7 | 2.2 | 3.0 |
| $R^{(2)}$ (Å) | 2.0 | 2.5 | 3.3 |
| $R^{(e)}$ (Å) | 1.3900 | 1.6345 | 2.0839 |

**Table 2.** Three-body parameters of the potential.

| Parameters | CCC | CCNi | CNiC | CNiNi | NiNiNi | NiNiC | NiCNi | NiCC |
|---|---|---|---|---|---|---|---|---|
| $\alpha$ (Å$^{-1}$) | 0 | 0 | 0 | 0 | 4.40 | 0 | 4.01 | 0 |
| $a$ | $2.08 \cdot 10^{-4}$ | 0.1 | 0.77 | $3.29 \cdot 10^{-3}$ | $9.28 \cdot 10^{-2}$ | 0 | $1.86 \cdot 10^{-4}$ | $1.22 \cdot 10^{-5}$ |
| $c$ | 330 | 0 | 0 | 5.72 | 7760 | 0 | 7410 | 240 |
| $d$ | 3.50 | 1.00 | 1.00 | 0.348 | 69.0 | 1.00 | 7.75 | 1.00 |

**Table 3.** Values of function $F_{CC}(i,j,k)$ for integer values of $i, j$ and $k$. Between integer values of $i, j$ and $k$, the function is interpolated by a cubic spline. All parameters not given are equal to zero, $F_{CC}(i,j,k) = F_{CC}(j,i,k)$, $F_{CC}(i,j,k>2) = F_{CC}(i,j,k)$.

|         | $F_{CC}$ |
|---------|----------|
| (1,1,1) | 0.1264   |
| (2,2,1) | 0.0605   |
| (1,2,1) | 0.012    |
| (1,3,1) | -0.0903  |
| (1,3,2) | -0.0903  |
| (0,3,1) | -0.0904  |
| (0,3,2) | -0.0904  |
| (0,2,2) | -0.0269  |
| (0,2,1) | 0.0427   |
| (0,1,1) | 0.0996   |
| (1,1,2) | 0.0108   |

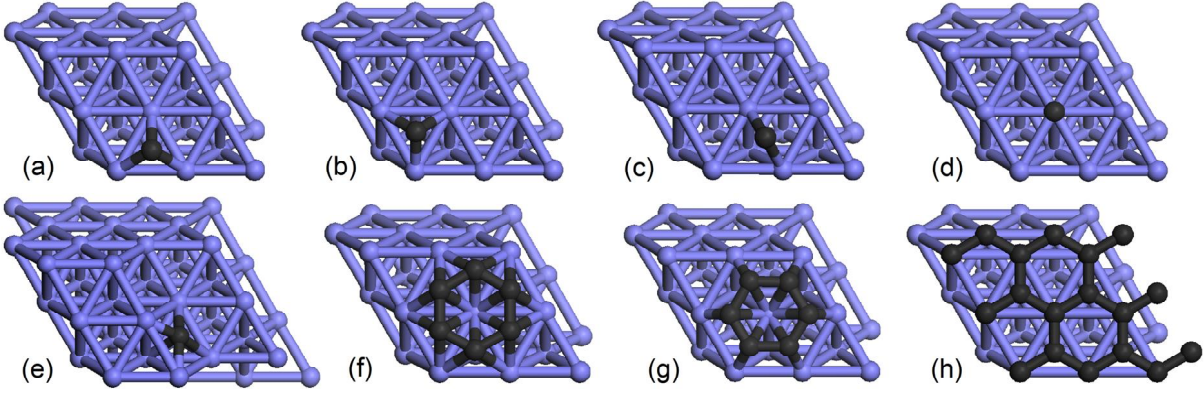

**Figure 1.** Structures of carbon species on the Ni (111) and (113) surfaces: carbon adatoms at (a) hcp, (b) fcc, (c) bridge, (d) top and (e) step sites; $C_6$ rings with atoms at (f) hcp/fcc and (g) bridge sites; (h) graphene.

**Table 4.** Structural and cohesive properties of Ni in different phases: lattice constant $a_0$, cohesive energy per atom $E_{coh}$, cohesive energy per atom relative to fcc phase $\Delta E_{coh}$, elastic constants $C_{11}$, $C_{12}$ and $C_{44}$, bulk modulus $B$, energy of vacancy formation $E_{vac}$, energy of the (111) surface $E_{(111)}$, energy of Ni adatom formation on the (111) surface $E_{adatom}$, energy of Ni addimer formation on the (111) surface $E_{addimer}$, energies of the (113) and (100) surfaces relative to the (111) surface $\Delta E_{(113)}$ and $\Delta E_{(100)}$, equilibrium distance of Ni dimer $R_e$, dissociation energy of Ni dimer $D_e$ and frequency of Ni dimer vibrations $v_0$. The energies are calculated for geometrically optimized structures except cases where the otherwise is stated.

| Parameter | Experiment | DFT | MEAM[17,23] | EAM[18] | Developed potential |
|---|---|---|---|---|---|
| Ni fcc | | | | | |
| $a_0$ (Å) | 3.52[a,b,c] | 3.53 | 3.52 | 3.52 | 3.52 |
| $E_{coh}$ (eV/atom) | -4.37[c], -4.44[b] | -4.76 | -4.45 | -4.43 | -4.37 |
| $C_{11}$ (GPa) | 246.5[d], 261.2[b] | | 252 | 231 | 241 |
| $C_{12}$ (GPa) | 147.3[d], 150.8[b] | | 146 | 177 | 150 |
| $C_{44}$ (GPa) | 124.7[d], 131.7[b] | | 127 | 79.6 | 132 |
| $B$ (GPa) | 180.4[d], 187.6[b] | 194.6 | 181 | 195 | 181 |
| $E_{vac}$ (eV) | 1.4[e] | | 1.46[f] | 1.11 (1.16[f]) | 1.08 (1.27[f]) |
| $E_{(111)}$ (J/m$^2$) | 2.24[g] | 1.91 | 2.04 | 1.53 | 1.52 |
| $E_{adatom}$ (eV) | | 1.14 | | | 1.25 |
| $E_{addimer}$ (eV) | | 1.92 | | | 2.13 |
| $\Delta E_{(113)}$ (J/m$^2$) | | 0.36 | | 0.193 | 0.367 |
| $\Delta E_{(100)}$ (J/m$^2$) | | | 0.40 | 0.103 | 0.384 |
| Ni bcc | | | | | |
| $a_0$ (Å) | | 2.81 | | 2.80 | 2.80 |
| $\Delta E_{coh}$ (eV/atom) | 0.07[h] | 0.096 | 0.09 | 0.033 | 0.024 |
| $B$ (GPa) | | 190 | | 181 | 192 |
| Ni hcp | | | | | |
| $\Delta E_{coh}$ (eV/atom) | 0.015[h] | | 0.021 | $-4.2 \cdot 10^{-5}$ | $8.5 \cdot 10^{-6}$ |
| Ni$_2$ | | | | | |

| | | | | |
|---|---|---|---|---|
| $R_e$ (Å) | 2.2[i] | 2.08 | 2.04 | 2.08 |
| $D_e$ (eV) | 2.068[i] | 2.70 | 4.23 | 2.70 |
| $\nu_0$, (cm$^{-1}$) | | 324 | 509 | 327 |

[a]Ref. 37; [b]Ref. 38; [c]Chemical Workbench database[39]; [d]Ref. 40; [e]Ref. 50; [f]for unrelaxed structures; [g]Ref. 41; [h]derived from the experimental stacking fault or phase diagram information (see Ref. 17 and references therein); [i]Ref. 49

**Table 5.** Calculated energies (in eV/atom) of carbon adatoms at different sites of the Ni (111) and (113) surfaces, of carbon atoms in $C_6$ and in graphene on the Ni (111) surface, and of carbon interstitials at the O (octahedral) and T (tetrahedral) sites in the subsurface layer and in the bulk relative to the energy of a carbon adatom at a hcp site of the Ni (111) surface. The energies are given for geometrically optimized structures. The structures of carbon species on the Ni surfaces are shown in Figure 1.

| Structure | DFT (present work) | DFT (previous works) | Developed potential | MEAM[23] |
|---|---|---|---|---|
| C at fcc site | 0.0561 | 0.05[a–c] | 0.0144 | |
| C at bridge site | 0.343 | 0.4,[c-e] 0.5[a,f] | 0.441 | |
| C at top site | 2.35 | | 2.19 | |
| C at step site | -0.98 | -0.32,[c] -0.7,[f] -1.0[a,e,g] | -0.757 | |
| $C_6$ with atoms at bridge sites | -0.326 | | -0.323 | |
| C₆ with atoms at hcp/fcp sites | -0.287 | | -0.284 | |
| graphene | -1.35 | -1.17,[b] -1.30,[a] | -1.42 | |
| | | -1.35[g] | | |
| C at subsurface O site | -0.472 | -0.34,[a] -0.42,[c] | -0.594 | |
| | | -0.5,[d] -0.54,[e] | | |
| | | -0.57[b] | | |
| C at bulk O site | -0.183 | -0.02,[a] -0.13,[e] | -0.128 | 3.11 |
| | | -0.16,[c] -0.2,[d] | | |
| | | -0.32[b] | | |
| C at bulk T site | 1.49 | 1.8,[a,c] 1.4[d] | 1.02 | 3.73 |

[a]Ref. 42; [b]Ref. 43; [c]Ref. 44; [d]Ref. 45; [e]Ref. 46; [f]Ref. 47; [g]Ref. 48

**2.3. Nickel-Nickel Interaction.** The parameters of the two-body terms of the potential for the Ni-Ni interaction (Table 1) are fitted to the characteristics of the nickel dimer obtained by the DFT calculations. The dissociation energy $D_e$, bond length $R_e$ and vibration frequency $\nu_0$ of the nickel dimer calculated using the developed potential are in excellent agreement with the results of the DFT calculations, though the bond length is somewhat lower and the dissociation energy is somewhat higher than the experimental data.[49] The parameters of the three-body terms of the potential for the Ni-Ni interaction (Table 2) are fitted to the experimental data on the lattice constant[37–39] $a_0$, cohesive energy[38,39] $E_{coh}$ and elastic properties[38,40] (elastic constants $C_{11}$, $C_{12}$ and $C_{44}$ and bulk modulus $B$) of fcc Ni. It is seen from Table 4 that the developed potential reproduces these quantities with an accuracy within 6%. The potential is qualitatively correct in predicting relative energies of Ni bulk phases (see Ref. 17 and references therein).

Though the absolute surface energy[41] $E_{(111)}$ of the (111) surface is underestimated by 30%, the relative energies $\Delta E_{(113)}$ and $\Delta E_{(100)}$ of the (113), (100) and (111) surfaces are reproduced within 4%. The energies $E_{adatom}$ and $E_{addimer}$ of formation of a Ni adatom and addimer on the (111) surface are overestimated by 10% compared to the results of our DFT calculations. The energy $E_{vac}$ of vacancy formation in fcc Ni is lower by 20% than the experimental value.[50]

The data taken from papers[17,23] for the MEAM potential and calculated by us using the Sutton-Chen EAM potential[18] are also shown in Table 4 for comparison. It is seen that the developed potential is more accurate than the commonly used Sutton-Chem EAM potential[18] (see the data for the elastic constants of fcc Ni and of the relative energies of Ni surfaces in Table 4). It should be also noted that the developed potential is somewhat less accurate than the MEAM potential[17,23] with regard to the absolute energies of the Ni surfaces and the relative energies of Ni phases. However, these properties are of minor importance for simulations of the effect of Ni clusters on the graphene-fullerene transformation.

**2.4. Nickel-Carbon Interaction.** The parameters for the Ni-C interaction (Table 1, Table 2) are fitted to the relative energies of carbon adatoms at different sites of the Ni (111) surface (at hcp, fcc, bridge and top sites), of carbon atoms in $C_6$ rings (with atoms at bridge sites and fcc/hcp sites) and in graphene on the Ni (111) surface and of carbon interstitials at O (octahedral) sites in the subsurface layer and in the bulk obtained in our DFT calculations. All of these quantities cannot be fitted simultaneously using the considered form of the potential. However, the fitted potential provides qualitatively correct relative energies of carbon atoms at different sites and in different carbon structures (Table 5). The energies of carbon atoms at bridge, top and O sites, in $C_6$ rings and graphene relative to carbon adatoms at hcp sites are reproduced with the accuracy within 30%. The relative energy of carbon adatoms at fcc and hcp sites of the Ni (111) surface is strongly underestimated. However, the potential gives the correct sign of this quantity. In addition to the above structures, the potential has been tested for the relative energies of carbon adatoms at steps of the Ni (113) surface and of carbon atoms in T (tetrahedral) sites in the Ni bulk. The energy of carbon adatoms at steps of the Ni (113) surface relative to carbon adatoms at hcp sites of the Ni (111) surface is underestimated by about 20%, while the barrier for carbon diffusion in the Ni bulk determined by the difference in the energies of carbon atoms at O and T sites is underestimated by more than 30% (Table 5).

It is also seen from Table 5 that the accuracy of the developed potential for calculation of relative energies of carbon atoms is much better than of the Ni-C MEAM potential[23] reported previously. The latter potential was fitted to the absolute energies of carbon adsorption on Ni surfaces, while the relative energies of carbon atoms in different positions and structures should be more relevant for description of growth of carbon nanostructures.

To summarize, we describe the Ni-Ni, Ni-C and C-C interactions using the same empirical potential which makes possible efficient simulations for Ni-C systems. The developed potential is shown to reproduce the physical properties of pure Ni and the binding energies of carbon adatoms, $C_6$ rings and graphene on the Ni (111) and (113) surfaces and of carbon interstitials in the Ni bulk and subsurface layer, while the original Brenner potential[19] is known to work perfectly for carbon structures. Thus the developed potential is optimum for simulations of catalyzed growth and transformations of carbon nanostructures. The same fitting procedure can be used to develop potentials for other metal-carbon systems.

## 3. MOLECULAR DYNAMICS SIMULATIONS

**3.1. Methodology.** The MD simulations have been performed for graphene flakes consisting of $N_C = 54$, 96, 216, 294 and 384 atoms and Ni clusters consisting of $N_{Ni} = 5$, 13 and 79 atoms. Initially the flakes have the shape of ideal hexagons with 6 equal zigzag edges (Figure 2a and Figure 3a). The in-house MD-kMC[51] (Molecular Dynamics – kinetic Monte Carlo) code is used. The integration time step is 0.6 ps. The simulations are performed at temperatures $T = 2700 - 3500$ K for free graphene flakes and at temperatures $T = 2000 - 3000$ K for graphene flakes interacting with Ni clusters. The temperature of the system is maintained by rescaling atomic velocities every 5 ps (the Berendsen thermostat[52]).

**3.2. Folding of Free Graphene Flakes.** Folding of free graphene flakes was previously investigated by us using MD simulations.[4,5] It was found that the folding occurs through transformation of hexagons at the edges of the bond network of the flake to various polygons. Principal reactions of formation of polygons different from hexagons were revealed, and parameters of these reactions were calculated. In the present paper, we perform the MD simulations of the graphene-fullerene transformation at different temperatures and derive the activation energy for this process. Free graphene flakes consisting of $N_C = 96$ and 384 atoms (with the length of each of 6 edges equal to 9.8 Å and 19.7 Å, respectively, see Figure 2a and Figure 3a) are considered as initial structures at temperatures $T = 2700 - 3500$ K. To clarify the mechanism of folding the total energy $E$ of the system and the number $N_2$ of two-coordinated atoms are monitored during the MD simulations (Figure 2 and Figure 3). To calculate the number $N_2$ we assume that two carbon atoms are bonded if the distance between them does not exceed 1.8 Å.

The graphene-fullerene transformation proceeds as follows (see supplemental movies in Supporting Information). At the beginning of the MD simulations bonds close to the flake edges get broken due to the thermal excitations, and chains of two-coordinated carbon atoms bound to the rest of the flake at their ends are formed (Figure 2b and Figure 3b). The formation of such chains is favorable as it is accompanied by a significant entropy increase.[4,5] So, first the number $N_2$ of two-coordinated atoms in the graphene flake increases (Figure 2f and Figure 3e). The total energy $E$ of the flake ascends correspondingly (Figure 2f and Figure 3e). Occasionally bonds between atoms at the flake edges are formed and broken again, leading to a graduate reconstruction of the bond topology. During a considerable period of time, the time-averaged number $N_2$ of two-coordinated atoms and the total energy $E$ of the flake stay nearly constant, while their instant values fluctuate by 20–30% relative to the initial value and by 0.05–0.1 eV/atom, respectively (Figure 2f and Figure 3e). As a result of bond breaking and formation, the graphene flake, which is almost flat initially, transforms at some moment into a bowl-shaped structure (Figure 2c and Figure 3c). For large graphene flakes, the formation of two bowl-shaped regions is possible (Figure 2d). In such bowl-shaped carbon clusters, relatively fast zipping of the edges occurs, leading to the drastic decrease of the number $N_2$ of two-coordinated atoms and the energy $E$ of the flake (Figure 2f and Figure 3e). Finally, a closed fullerene structure is formed (Figure 2e and Figure 3d).

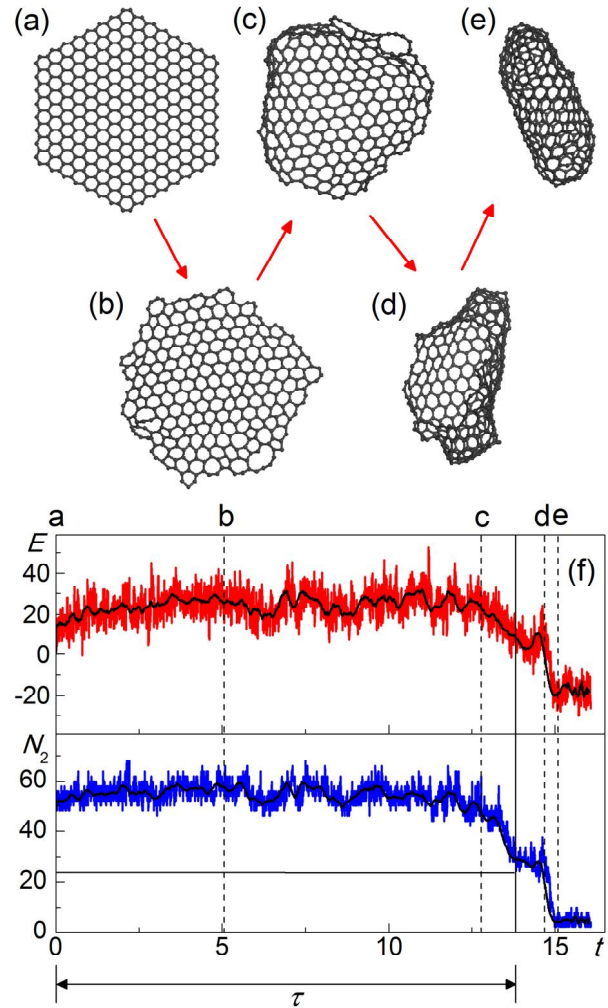

**Figure 2.** (a – e) Structures obtained at different moments of time in the MD simulations of the transformation of the $C_{384}$ graphene flake at temperature 3000 K: (a) 0 ns, (b) 5.1 ns, (c) 12.8 ns, (d) 14.7 ns and (e) 15.1 ns. (f) Calculated total energy $E$ of this system (in eV; red line) and number $N_2$ of two-coordinated atoms (blue line) as functions of time $t$ (in ns). The energy is given relative to the initial value. Moments of time corresponding to structures (a – e) are shown using dashed lines. The folding time $\tau$ is indicated by straight solid lines and a double-headed arrow. Black solid curves are shown to guide the eye.

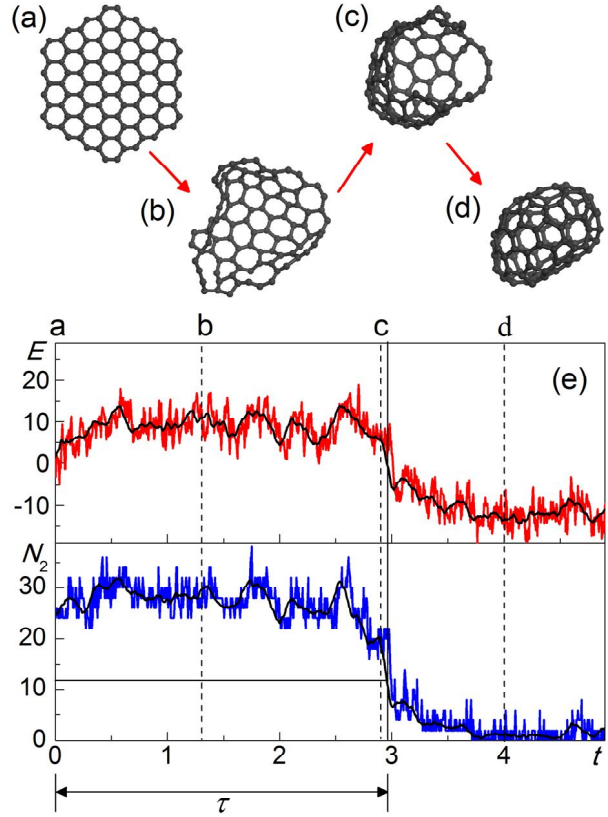

**Figure 3.** (a – d) Structures obtained at different moments of time in the MD simulations of the transformation of the $C_{96}$ graphene flake at temperature 3000 K: (a) 0 ns, (b) 1.3 ns, (c) 2.9 ns and (d) 4.0 ns. (e) Calculated total energy $E$ of this system (in eV; red line) and number $N_2$ of two-coordinated atoms (blue line) as functions of time $t$ (in ns). The energy is given relative to the initial value. Moments of time corresponding to structures (a – d) are shown using dashed lines. The folding time $\tau$ is indicated by straight solid lines and a double-headed arrow. Black solid curves are shown to guide the eye.

To characterize the kinetics of the graphene-fullerene transformation in our simulations we introduce the folding time $\tau$. We determine this time as the moment when the number $N_2$ of two-coordinated atoms reaches half of the initial value $N_2(\tau) = N_2(0)/2$ (Figure 2f and Figure 3e). The average folding times $\langle \tau \rangle$ calculated for the $C_{96}$ and $C_{384}$ graphene flakes at different temperatures are given in Table 6.

**Table 6.** Calculated average folding times $\langle\tau\rangle$ and root-mean-square deviations $\sigma$ of these times for different temperatures $T$ and sizes $N_\text{C}$ of free graphene flakes. The total number $N$ of simulations for each considered system is indicated.

| $N_\text{C}$ | $T$ (K) | $N$ | $\langle\tau\rangle$ (ns) | $\sigma$ (ns) |
|---|---|---|---|---|
| 384 | 3500 | 8 | 1.5 | 0.46 |
| 384 | 3200 | 8 | 6.7 | 1.7 |
| 384 | 3000 | 7 | 16 | 4.1 |
| 96 | 3000 | 30 | 4.6 | 2.8 |
| 96 | 2800 | 30 | 16 | 8.6 |
| 96 | 2700 | 30 | 25 | 14 |

It is seen that the average folding times $\langle\tau\rangle$ decrease with increasing temperature $T$ and decreasing the flake size $N_\text{C}$. We approximate the average folding times $\langle\tau\rangle$ for the considered flakes with a simple Arrhenius expression

$$\langle\tau\rangle = \tau_0 \, exp\left(\frac{E_\text{a}}{k_\text{B}T}\right), \tag{12}$$

where $\tau_0$ is the pre-exponential factor, $E_\text{a}$ is the activation energy for folding and $k_\text{B}$ is the Boltzmann constant. Figure 4 demonstrates that the use of Arrhenius expression (12) is adequate for description of the dependence of the average folding time on temperature. The activation energies $E_\text{a}$ and the pre-exponential factors $\tau_0$ estimated for the $C_{96}$ and $C_{384}$ graphene flakes using the data from Table 6 are given in Table 7. It is seen from Table 7 that the activation energies are about $E_\text{a} \sim 4$ eV for both the considered flakes. The difference in the activation energies for the flakes is smaller than the calculation error. However, the average folding time $\langle \tau \rangle$ for the $C_{384}$ flake at temperature 2500 K exceeds that for the $C_{96}$ flake by a factor of 3 – 4 (Table 6). Using Eq. (12) and based on the activation energies $E_a$ and the pre-exponential factors $\tau_0$ obtained for the free graphene flakes (Table 7), we estimate that folding of such graphene flakes can be observed experimentally at times $\langle \tau \rangle \sim 1$ s at temperatures above 1400 K.

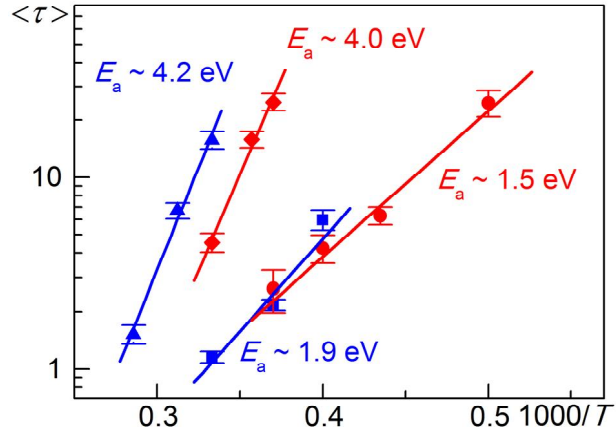

**Figure 4.** Calculated average times $\langle \tau \rangle$ of folding of graphene flakes (in ns) as functions of the reciprocal of temperature $1000/T$ (in K$^{-1}$) for the free $C_{96}$ flake (red diamonds), for the $C_{96}$ flake with the $Ni_{13}$ cluster attached (red circles), for the free $C_{384}$ flake (blue triangles) and for the $C_{384}$ flake with the $Ni_{79}$ cluster attached (blue squares). The solid lines show Arrhenius approximations (12) of the obtained dependences. The estimated activation energies are indicated.

**3.3. Influence of Ni Clusters on Folding of Graphene Flakes.** To study the effect of Ni clusters on the graphene-fullerene transformation we have performed MD simulations for graphene flakes consisting of $N_C = 54, 96, 216, 294$ and 384 atoms (with the length of each of 6 edges equal to 7.4, 9.8, 14.8, 17.2 and 19.7 Å, respectively) and Ni clusters consisting of $N_{Ni} = 5, 13$ and 79 atoms. Initially a Ni cluster is placed at a corner of a graphene flake (Figure 5a, Figure 6a and Figure 7a) so that strong Ni-C bonds are formed. The acceleration of kinetics of the graphene-fullerene transformation in the presence of Ni clusters allows decreasing temperatures in the MD simulations down to 2000 – 3000 K as compared to the simulations of the free flakes at temperatures 2700 – 3500 K.

To characterize the Ni-C system during the MD simulations, we distinguish several groups of carbon atoms. The network of C-C bonds is analyzed. Carbon atoms which are not bound by C-C bonds to the flake (i.e., are isolated from the flake) or chains of carbon atoms which are bound to the flake at only one end are considered as dissolved in the Ni cluster or adsorbed on the cluster surface (such types of carbon atoms are very unstable unless they are attached to the Ni cluster and usually are not observed for free flakes). The number of such atoms is denoted as $N_d$. Then the network formed by the rest of carbon atoms is analyzed. In this network the number $N_2'$ of effective "two-coordinated" carbon atoms having two bonds with atoms in the network is determined. The number $N_2'$ characterizes the perimeter of the graphene flake, and a considerable decrease of this number corresponds to the formation of a fullerene shell. Among all effective "two-coordinated" carbon atoms, those which are bound to the cluster are then counted. The number $N_{CNi}$ of these effective "two-coordinated" carbon atoms bound to the Ni cluster is calculated under the assumption that carbon and nickel atoms are bonded if the distance between them does not exceed 2.2 Å. The number $N_{CNi}$ characterizes the contact length between the flake and cluster. A significant increase of this number relative to the initial value corresponds to wrapping the flake around the cluster. The dependences of numbers $N_d$, $N_2'$ and $N_{CNi}$ on time obtained using the MD simulations help us to investigate the mechanism and kinetics of the Ni-assisted graphene-fullerene transformation.

**Table 7.** Estimated activation energies $E_a$ and pre-exponential factors $\tau_0$ for folding of graphene flakes of different size $N_C$ without Ni clusters and attached to Ni clusters of size $N_{Ni}$.

| $N_C$ | $N_{Ni}$ | $E_a$ (eV) | $\tau_0$ (s) |
|---|---|---|---|
| 384 | - | 4.2 ± 0.3 | $10^{-14.9 \pm 0.5}$ |
| 96 | - | 4.0 ± 0.4 | $10^{-15.0 \pm 0.6}$ |
| 384 | 79 | 1.9 ± 0.2 | $10^{-12.2 \pm 0.3}$ |
| 96 | 13 | 1.5 ± 0.2 | $10^{-11.5 \pm 0.4}$ |

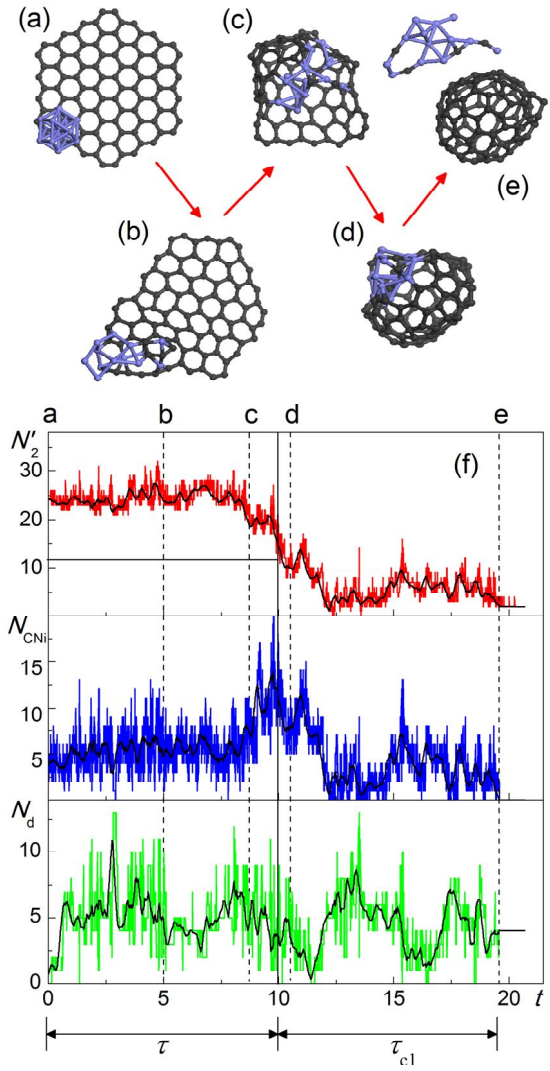

**Figure 5.** (a – e) Structures obtained at different moments of time in the MD simulations of the transformation of the $C_{96}$ graphene flake with the $Ni_{13}$ cluster attached at temperature 2300 K: (a) 0 ns, (b) 5.0 ns, (c) 8.7 ns, (d) 10.5 ns and (e) 19.6 ns. (f) Calculated number $N'_2$ of effective "two-coordinated" carbon atoms (red line), $N_{CNi}$ of effective "two-coordinated" carbon atoms bound to the Ni cluster (blue line) and number $N_d$ of carbon atoms dissolved in the Ni cluster and adsorbed on the cluster surface (green line) as functions of time $t$ (in ns) for this system. Moments of time corresponding to structures (a – e) are shown by dashed lines. The folding time $\tau$ and the time $\tau_{cl}$ of closure of the fullerene shell are indicated by straight solid lines and double-headed arrows. Black solid curves are shown to guide the eye.

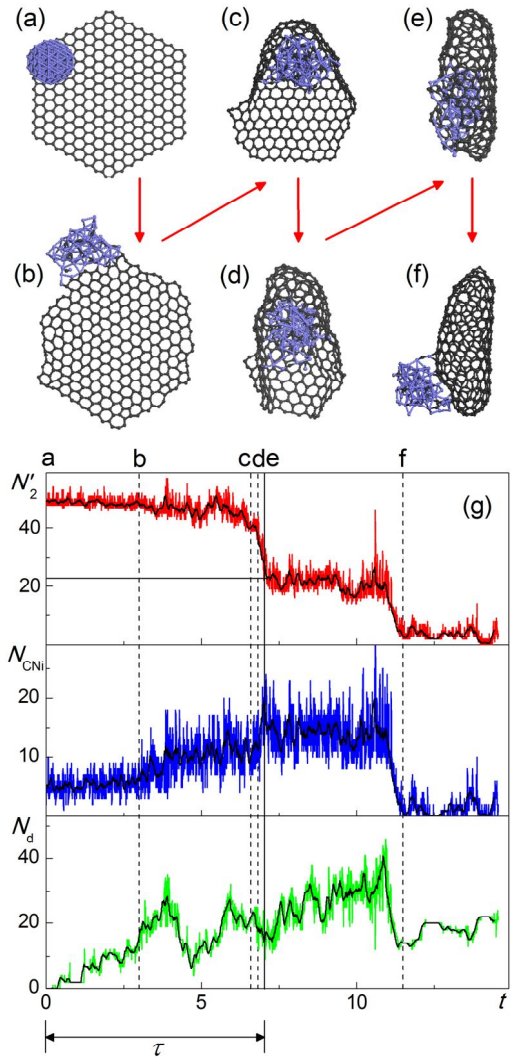

**Figure 6.** (a – f) Structures obtained at different moments of time in the MD simulations of transformation of the $C_{384}$ graphene flake with the $Ni_{79}$ cluster attached at temperature 2500 K: (a) 0 ns, (b) 3.0 ns, (c) 6.6 ns, (d) 6.8 ns, (e) 7.0 ns and (f) 11.5 ns. (g) Calculated number $N'_2$ of effective "two-coordinated" carbon atoms (red line), $N_{CNi}$ of effective "two-coordinated" carbon atoms bound to the Ni cluster (blue line) and number $N_d$ of carbon atoms dissolved in the Ni cluster and adsorbed on the cluster surface (green line) as functions of time $t$ (in ns) for this system. Moments of time corresponding to structures (a – f) are shown by dashed lines. The folding time $\tau$ is indicated by straight solid lines and a double-headed arrow. Black solid curves are shown to guide the eye.

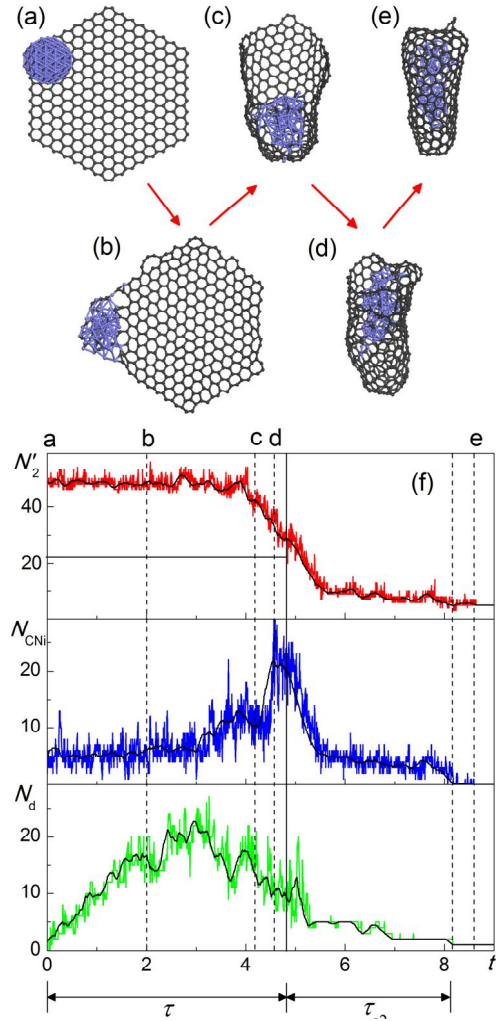

**Figure 7.** (a – e) Structures obtained at different moments of time in the MD simulations of transformation of the $C_{384}$ graphene flake with the $Ni_{79}$ cluster attached at temperature 2500 K: (a) 0 ns, (b) 2.0 ns, (c) 4.2 ns, (d) 4.6 ns and (e) 8.6 ns. (f) Calculated number $N'_2$ of effective "two-coordinated" carbon atoms (red line), $N_{CNi}$ of effective "two-coordinated" carbon atoms bound to the Ni cluster (blue line) and number $N_d$ of carbon atoms dissolved in the Ni cluster and adsorbed on the cluster surface (green line) as functions of time $t$ (in ns) for this system. Moments of time corresponding to structures (a – e) are shown by dashed lines. The folding time $\tau$ and the time $\tau_{c2}$ of closure of the fullerene shell are indicated by straight solid lines and double-headed arrows. Black solid curves are shown to guide the eye.

The MD simulations reveal that folding of the graphene flakes in the presence of the Ni clusters (Figure 5a, Figure 6a and Figure 7a) proceeds as follows (see supplemental movies in Supporting Information). First, the cluster diffuses along the flake edges (Figure 5b, Figure 6b and Figure 7b). The formation of polygons different from hexagons is observed at the edges of the bond network of the flake both in the contact area between the graphene flake and the Ni cluster and out of this area. However, the formation of chains of two-coordinated atoms, which are responsible for folding of free graphene flakes, is relatively rare at lower temperatures considered in the simulations of the Ni-assisted transformation. This is supported by the fact that the number $N'_2$ of effective "two-coordinated" carbon atoms does not increase much with respect to the initial value (Figure 5f, Figure 6g and Figure 7f). Instead, many reactions in the contact area between the graphene flake and the Ni cluster occur with the detachment of carbon atoms from the flake followed by their diffusion into the Ni cluster and on the cluster surface. It is seen from Figure 5f, Figure 6g and Figure 7f that the number $N_d$ of dissolved and adsorbed carbon atoms increases and reaches up to half of the number $N_{Ni}$ of nickel atoms. At some moment, carbon atoms dissolved in the Ni cluster or adsorbed on the cluster surface start attaching back to the flake and condensate into a reconstructed part of the flake with an increased number of bonds to the cluster. Thus, a decrease of the number $N_d$ of carbon atoms dissolved in the Ni cluster and adsorbed on the cluster surface is accompanied by an increase of the number $N_{CNi}$ of "two-coordinated" carbon atoms bound to the Ni cluster (Figure 5f, Figure 6g and Figure 7f). The transfer of carbon atoms from the graphene flake to the Ni cluster and back accompanied by reconstruction of the bond network of the flake takes place until the flake transforms into a bowl-shaped structure and wraps around the cluster (Figure 5c, Figure 6c and Figure 7c). As this happens, the number of $N_{CNi}$ of "two-coordinated" carbon atoms bound to the Ni cluster reaches its maximum, while the number $N'_2$ of effective "two-coordinated" carbon atoms starts decreasing (Figure 5f, Figure 6g and Figure 7f). The cluster moves along the flake edges bringing them together and zipping (Figure 6d and Figure 7d). During this time, the number $N_{CNi}$ of effective "two-coordinated" carbon atoms bound to the Ni cluster remains large, while the number $N'_2$ of effective

"two-coordinated" carbon atoms decreases drastically (Figure 5f, Figure 6g and Figure 7f). When the cluster reaches the end of the flake edges, a fullerene shell forms. Then the cluster can either stay inside of the fullerene shell and provide its complete closure (Figure 7e) or form a patch to the fullerene shell (Figure 5d, Figure 6e). In the latter case, the cluster can get in and out (Figure 6f) of the fullerene shell several times and finally detach from the fullerene (Figure 5e). As the fullerene shell is completely closed, the number $N_{\mathrm{CNi}}$ of effective "two-coordinated" carbon atoms bound to the Ni cluster goes to zero and the number $N'_2$ of effective "two-coordinated" carbon atoms becomes negligibly small (Figure 5f and Figure 7f).

Therefore, the Ni-assisted graphene-fullerene transformation proceeds differently from the transformation of free graphene flakes and the transfer of carbon atoms from the graphene flake to the Ni cluster and back plays a crucial role in this process. The visual analysis of the evolution of the system structure with time reveals that pairs (there are initially six such pairs in hexagons at the flake corners, see Figure 5a, Figure 6a and Figure 7a) or longer chains of two-coordinated atoms can detach from the flake at one of their ends and dissociate into carbon atoms, which further diffuse into the Ni cluster or on the cluster surface, while isolated two-coordinated carbon atoms or three-coordinated carbon atoms of the flake do not detach (see supplemental movies in Supporting Information). The condensation of carbon atoms from the Ni cluster leads to an addition of hexagons and rarely of pentagons to the bond network of the flake. Based on these observations, we suggest a possible pathway (Figure 8) of formation of an extra pentagon surrounded by hexagons in the bond network of the flake. Hexagons at the flake edges easily transform to other polygons.[4,5] For example, a pentagon and a heptagon can be formed from two adjacent hexagons (Figure 8a and b). The pair of two-coordinated carbon atoms can then detach from the flake providing elimination of the heptagon (Figure 8c) and dissociate into carbon atoms, which diffuse into the Ni cluster or on the cluster surface (Figure 8d). One of carbon atoms dissolved in the Ni cluster or adsorbed on the cluster surface then fills the void, forming a hexagon (Figure 8d and e). Thus, an excess of pentagons over large polygons (*i.e,.* heptagons in this example) arises. An addition of another pair of carbon atoms to the flake results in the creation of a group of polygons consisting of the pentagon surrounded by hexagons (Figure 8f). Formation of several pentagons surrounded by hexagons according to this pathway leads to the transformation of the flake to the bowl-shaped structure.

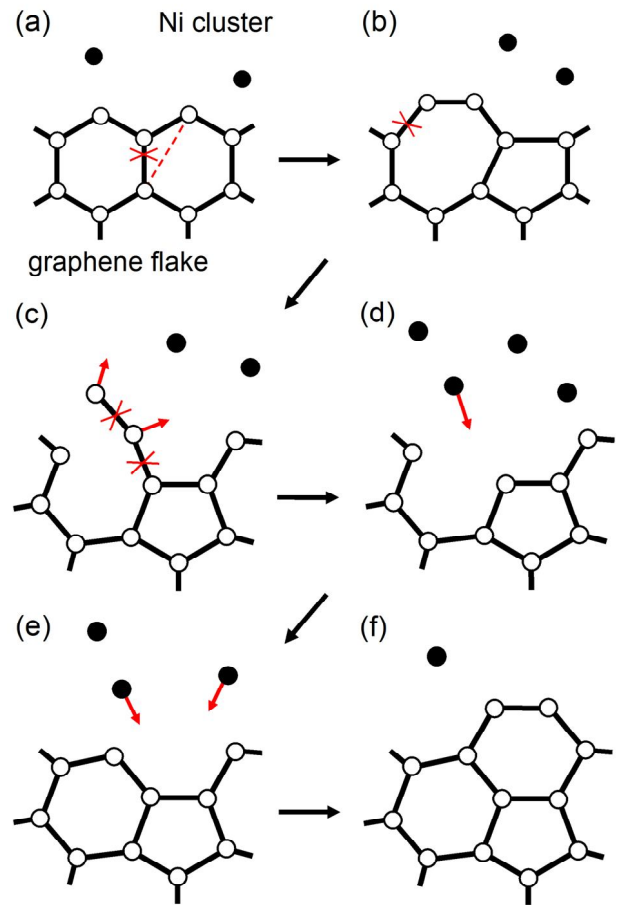

**Figure 8.** Schematic representation of the mechanism of formation of a pentagon surrounded by hexagons in the bond network of the graphene flake in the presence of the Ni cluster. Carbon atoms in the graphene flake are shown with open circles. Carbon atoms dissolved in the Ni cluster or adsorbed on the cluster surface are shown with filled circles. Bonds between carbon atoms are shown with black solid lines. Bond breaking and formation is indicated with red crosses and dashed lines. The detachment of carbon atoms from the graphene flake followed by their diffusion into the Ni cluster or on the cluster surface and attachment of carbon atoms back to the graphene flake are indicated with red arrows.

Three types of final structures are found in the MD simulations of the Ni-assisted graphene-fullerene transformation: 1) a free fullerene (Figure 5e), 2) a fullerene with a Ni cluster attached from outside (Figure 6f), and 3) a fullerene encapsulating a Ni cluster (Ni endofullerene, Figure 7e). In the first case, the Ni cluster completely gets out of the fullerene shell formed in the result of folding of the graphene flake and detaches from the fullerene (Figure 5e). This is usually observed for the small graphene flakes consisting of $N_C = 54 - 216$ atoms and the Ni clusters consisting of $N_{Ni} = 5$ and 13 atoms. For the graphene flakes consisting of $N_C = 294$ and 384 atoms and the Ni clusters consisting of $N_{Ni} = 13$ and 79 atoms, the Ni cluster is often seen to be attached to the fullerene from outside (Figure 6f). We believe that at the considered temperatures the Ni cluster should finally detach from the fullerene and fly away, but the detachment times for these systems are beyond the times accessible for our MD simulations. For the graphene flakes consisting of $N_C = 216 - 384$ atoms and the Ni clusters consisting of $N_{Ni} = 13$ and 79 atoms, formation of Ni endofullerenes is also observed (Figure 7e). The fractions $\xi$ of simulations in which Ni endofullerenes are formed are given in Table 8. It is interesting to note that a new type of heterofullerenes, namely a heterofullerene with a patch made of the Ni cluster, is formed as an intermediate structure during the graphene-fullerene transformation. The possibility to obtain such heterofullerenes as a final result of the transformation is discussed in the Conclusion.

Based on the dependence of the number $N'_2$ of effective "two-coordinated" carbon atoms on time, the folding time $\tau$ is calculated in the same way as previously for folding of free graphene flakes, i.e., as the moment of time when the number $N'_2$ decreases by a factor of 2 (Figure 5f, Figure 6g and Figure 7f). The average folding times $\langle \tau \rangle$ obtained are listed in Table 8. From comparison of Table 6 and Table 8, it is clearly seen that the average folding times $\langle \tau \rangle$ for the $C_{384}$ flake at temperature $T = 3000$ K and for $C_{96}$ flake at temperature $T = 2700$ K decrease in the presence of the $Ni_{79}$ and $Ni_{13}$ clusters, respectively, by an order of magnitude. It also follows from Table 8 that the average folding time $\langle \tau \rangle$ decreases with increasing temperature $T$ (see the data for $N_{Ni} = 79$, $N_C = 384$; $N_{Ni} = 13$, $N_C = 96$; $N_{Ni} = 5$, $N_C = 54$), decreasing the flake size $N_\text{C}$ (see the data for $N_\text{Ni}=13$, $T=2500$ K) or increasing the Ni cluster size $N_\text{Ni}$ (compare the data for $N_\text{Ni}=13$, $N_\text{C}=216$, 294, $T=2500$ K and $N_\text{Ni}=79$, $N_\text{C}=384$, $T=2500$ K). The same as for the free graphene flakes, we use Arrhenius expression (12) to approximate the dependences of the average folding times $\langle\tau\rangle$ on temperature $T$ for the $C_{96}$ flake with the $Ni_{13}$ cluster and for the $C_{384}$ flake with the $Ni_{79}$ cluster. Figure 4 demonstrates that such an approach is adequate for approximation of the results on the Ni-assisted graphene-fullerene transformation. The estimated activation energies and pre-exponential factors are given in Table 7. It is seen from Table 7 that both for the $C_{96}$ and $C_{384}$ graphene flakes the Ni clusters provide a decrease of the activation energy by more than a factor of 2 compared to the free flakes. It should also be mentioned that different from the case of the free graphene flakes, for which the activation energy weakly depends on the flake size (Table 7), the activation energy for the Ni-assisted graphene-fullerene transformation depends on the system under consideration. The activation energy for the $C_{96}$ flake with the $Ni_{13}$ cluster is found to be smaller than the activation energy for the $C_{384}$ flake with the $Ni_{79}$ cluster by 0.4 eV. Using Eq. (12) and based on the activation energies $E_\text{a}$ and the pre-exponential factors $\tau_0$ obtained for the $C_{96}$ flake with the $Ni_{13}$ cluster and for the $C_{384}$ flake with the $Ni_{79}$ cluster (Table 7), we estimate that the Ni-assisted graphene-fullerene transformation can be observed in these systems experimentally at times $\langle\tau\rangle\sim1$ s at temperatures above 700 – 800 K. Thus, the minimal temperature at which the graphene-fullerene transformation can be observed experimentally is significantly lower in the presence of Ni clusters compared to the case of free graphene flakes.

**Table 8.** Calculated average folding times $\langle\tau\rangle$ and times $\langle\tau_{c1}\rangle$ and $\langle\tau_{c2}\rangle$ of closure of the fullerene shell, root-mean-square deviations $\sigma$, $\sigma_{c1}$ and $\sigma_{c2}$ of these times, fractions $\chi$ of simulations in which the Ni cluster detaches from the graphene flake before the folding occurs and fractions $\xi$ of simulations in which Ni endofullerenes are formed for different temperatures $T$, sizes $N_C$ of graphene flakes and sizes $N_{Ni}$ of Ni clusters attached. The total number $N$ of simulations for each considered system is indicated. The folding times $\tau$ and times $\tau_{c1}$ and $\tau_{c2}$ of closure of the fullerene shell are averaged over the simulations in which the Ni cluster stays attached to the flake during the folding. The average times $\langle\tau_{c1}\rangle$ of closure of the fullerene shell are calculated for simulations in which free fullerenes are formed. The average times $\langle\tau_{c2}\rangle$ of closure of the fullerene shell are calculated for simulations in which Ni endofullerenes are formed.

| $N_C$ | $N_{Ni}$ | $T$ (K) | $N$ | $\chi$ | $\xi$ | $\langle\tau\rangle$ (ns) | $\sigma$ (ns) | $\langle\tau_{c1}\rangle$ (ns) | $\sigma_{c1}$ (ns) | $\langle\tau_{c2}\rangle$ (ns) | $\sigma_{c2}$ (ns) |
|---|---|---|---|---|---|---|---|---|---|---|---|
| 384 | 79 | 2500 | 8 | 1/8 | 4/8 | 6.0 | 1.9 | | | 2.1 | 1.1 |
| 384 | 79 | 2700 | 12 | 0/12 | 7/12 | 2.2 | 0.49 | | | 3.1 | 1.1 |
| 384 | 79 | 3000 | 11 | 0/12 | 7/12 | 1.2 | 0.26 | | | 1.6 | 0.86 |
| 96 | 13 | 2000 | 8 | 1/8 | 0/8 | 25 | 10 | > 55 | | | |
| 96 | 13 | 2300 | 15 | 1/15 | 0/15 | 6.3 | 2.5 | 22 | 19 | | |
| 96 | 13 | 2500 | 15 | 0/15 | 0/15 | 4.2 | 2.7 | 8.5 | 6.9 | | |
| 96 | 13 | 2700 | 15 | 0/15 | 0/15 | 2.6 | 2.5 | 3.5 | 3.3 | | |
| 216 | 13 | 2500 | 15 | 4/15 | 1/15 | 19.0 | 6.6 | 3.2 | 2.7 | ~3 | |
| 294 | 13 | 2500 | 15 | 12/15 | 1/15 | 26 | 8.1 | 1.8 | 1 | ~1 | |
| 54 | 5 | 2500 | 15 | 10/15 | 0/15 | 5.5 | 1.9 | 7.6 | 6.6 | | |
| 54 | 5 | 2300 | 15 | 3/15 | 0/15 | 29 | 13 | 19 | 14 | | |

| | | | | | | | |
|---|---|---|---|---|---|---|---|
| 96 | 5 | 2300 | 15 | 14/15 | 0/15 | ~20 | ~20 |

For the simulations in which the Ni cluster detaches from the fullerene or a Ni endofullerene is formed, we also estimate the times $\tau_{c1}$ and $\tau_{c2}$ of closure of the fullerene shell as the difference between the time when the number $N'_2$ of effective "two-coordinated" carbon atoms becomes stationary and the folding time $\tau$ (Figure 5f and Figure 7f). The average times $\langle \tau_{c1} \rangle$ and $\langle \tau_{c2} \rangle$ of closure of the fullerene shell for these systems are listed in Table 8. From the data for the $C_{96}$ flake with the $Ni_{13}$ cluster and for the $C_{54}$ flake with the $Ni_5$ cluster, it is seen that the average time $\langle \tau_{c1} \rangle$ of closure of the fullerene shell in the case when the Ni cluster detaches from the fullerene strongly decreases with increasing temperature. Using a simple Arrhenius expression for the average time $\langle \tau_{c1} \rangle$ similar to Eq. (12) for the average folding time $\langle \tau \rangle$, we estimate the activation energies for the detachment of the $Ni_{13}$ and $Ni_5$ clusters from the fullerenes to be about 2.5 eV and 2.3 eV, respectively. The average time $\langle \tau_{c1} \rangle$ is also found to increase with decreasing the size of the graphene flake (see Table 8 for $N_{Ni} = 13$, $T = 2500$ K). The average time $\langle \tau_{c2} \rangle$ of closure of the fullerene shell in the case when the Ni cluster is encapsulated inside the fullerene is observed to weakly depend on temperature (see Table 8 for $N_{Ni} = 79$ and $N_C = 384$). When the Ni cluster does not leave the fullerene within the simulation time and remains attached to the fullerene shell from outside, the fullerene shell is seen to be broken now and then, and the number $N_2'$ of effective "two-coordinated" carbon atoms is not stationary (Figure 6g). Thus, the time of closure of the fullerene shell in this case cannot be evaluated.

Though the Ni clusters clearly accelerate the graphene-fullerene transformation, we observe that in some simulations the cluster detaches from the graphene flake before it folds. The fractions $\chi$ of simulations in which the Ni cluster detaches from the graphene flake before the folding occurs for the flakes and Ni clusters of different size at different temperatures are given in Table 8. It is seen that the fraction $\chi$ of such simulations is high for the large flakes and small clusters at high temperatures. This fraction decreases with decreasing the flake size $N_C$ (see Table 8 for $N_{Ni} = 5$, $T = 2300$ K and $N_{Ni} = 13$, $T = 2500$ K) or temperature $T$ (see Table 8 for $N_{Ni} = 5$ and $N_C = 54$). The energy $E_d$ required for the cluster detachment can be estimated as $E_d \approx N_{CNi}\varepsilon_{CNi}$, where $\varepsilon_{CNi}$ is the binding energy of two-coordinated carbon atoms at the flake edges to the Ni cluster. We calculate this energy $\varepsilon_{CNi}$ for the Ni (111) surface to be $\varepsilon_{CNi} \approx 0.9$ eV/atom. The initial values of $N_{CNi}$ for the clusters with $N_{Ni} = 5 - 79$ lie in the range of $3 - 5$. Thus the energy $E_d$ required for the detachment of these clusters is estimated to be $2.7 - 4.5$ eV, which is considerably higher than the calculated activation energies for folding of the $C_{96}$ flake with the $Ni_{13}$ cluster and for the $C_{384}$ flake with the $Ni_{79}$ cluster (Table 7). Therefore, with decreasing temperature, the fraction of cases in which the detachment of Ni clusters occurs faster than folding of graphene flakes should become negligibly small.

## 4. CONCLUSION

We have extended the Brenner potential[19] to describe Ni-C systems. The potential reproduces the experimental data on the lattice constant, cohesive energy, elastic properties and energy of vacancy formation for fcc Ni and is qualitatively correct in predicting relative stability of Ni bulk phases. The relative energies of the Ni surfaces and the energies of formation of a Ni adatom and addimer on the (111) surface obtained by the DFT calculations are reproduced by the potential. The potential also correctly describes the relative energies of carbon adatoms at different sites of the Ni (111) and (113) surfaces, of atoms in $C_6$ rings and in graphene on the Ni (111) surface and of carbon interstitials in the subsurface layer and in the bulk obtained by the DFT calculations. Therefore, the developed potential is suitable for simulations of catalyzed growth and transformations of carbon nanostructures.

The potential has been applied for the MD simulations of the graphene-fullerene transformation for graphene flakes with and without Ni clusters attached. Folding of free graphene flakes is shown to proceed through formation of chains of two-coordinated carbon atoms at the flake edges. The activation energy for this process is calculated to be about 4 eV for the flakes consisting of $100 - 400$ atoms. The average folding times are found to decrease with decreasing the flake size. The minimal temperature at which the transformation of free graphene flakes to fullerenes can be observed experimentally is estimated to be about 1400 K.

The MD simulations of the Ni-assisted graphene-fullerene transformation show that the transfer of carbon atoms from the graphene flake to the Ni cluster and back plays a crucial role in this process.

Based on the analysis of the system structure during the MD simulations we have revealed a possible pathway of formation of an excess of pentagons over heptagons and larger polygons in the bond network of the flake which eventually provides folding of the initially flat graphene flake to the fullerene. The proposed pathway includes such steps as (1) formation of polygons different from hexagons (*e.g.*, heptagon-pentagon pairs) at the edges of the bond network of the graphene flake, (2) detachment of pairs or chains of two-coordinated carbon atoms from heptagons or larger polygons of the flake followed by their dissociation into carbon atoms, which diffuse into the Ni cluster or on the cluster surface, and (3) attachment of dissolved or adsorbed carbon atoms back to the graphene flake leading to formation of hexagons. The activation energies for folding of the $C_{96}$ flake with the $Ni_{13}$ cluster attached and for the $C_{384}$ flake with the $Ni_{79}$ cluster attached are calculated to be 1.5 and 1.9 eV, respectively, which are more than twice smaller than the activation energy for folding of the free graphene flakes. The average folding times are shown to decrease with decreasing the flake size and increasing the Ni cluster size. The minimal temperature at which the Ni-assisted graphene-fullerene transformation can be observed experimentally is estimated to be 700 – 800 K.

Three types of final structures are obtained in the MD simulations of the Ni-assisted graphene-fullerene transformation: (1) a free fullerene, (2) a fullerene with a Ni cluster attached from outside, and (3) a fullerene encapsulating a Ni cluster (a Ni endofullerene). Ni-endofullerenes are formed in 50–60% of simulations for the $C_{384}$ flake with the $Ni_{79}$ cluster at temperatures 2500–3000 K. However, such a final structure is rarely observed for the Ni cluster consisting of 13 atoms at temperatures 2000 – 2700 K. The Ni cluster consisting of 5 atoms is always found to fly out of the fullerenes at temperatures 2300 – 2500 K. The analogous escape of transition metal atoms was observed at heating double-walled carbon nanotubes formed by a catalytic process from single-walled carbon nanotube filled with ferrocenes.[13] Such an escape can explain the absence of endofullerenes and heterofullerenes which contain atoms of transition metals (See Refs. 53 and 54 for reviews with the list of observed endofullerenes and heterofullerenes).

Moreover, a new type of heterofullerenes, namely a heterofullerene with a patch made of a Ni cluster, is observed as an intermediate structure in the MD simulations of the Ni-assisted graphene-fullerene transformation. We propose that folding of graphene flakes with transition metal clusters attached can also take place at room temperature under the action of an electron beam in the transmission electron microscope (analogously to folding of free graphene flakes[3]). Such a transformation could be interrupted at the moment when the cluster is built into the fullerene shell as a patch or when the cluster gets into the fullerene shell and thus can be used to for a controlled synthesis of heterofullerenes with a transition metal patch or endofullerenes with transition metal clusters inside.

Since pioneering works[55,56] of Eigler a considerable progress has been achieved in controlled manipulation of atoms and molecules on surfaces (see reviews for atom/molecule manipulation using scanning tunneling microscopy[57] and atomic force microscopy[58]). For example, metal nanoclusters on the surface were assembled and disassembled with a precise control of single atoms.[59] The methods of manipulation of graphene flakes have been also developed: small graphene flakes were moved on a graphite surface by the tip of the friction force microscope,[60,61] pull-out of graphite flakes from graphite stacks[62] and cutting a graphene layer into flakes of certain size and shape[15] were demonstrated. Thus, recent technologies make possible preparation of metal clusters consisting of a controlled number of atoms (atoms of different elements can be combined into one cluster) attached to graphene flakes of a controlled shape. These give us a cause for optimism that both the temperature-activated transition metal-assisted graphene-fullerene transformation and the proposed way of the controlled synthesis of new types of endofullerenes and heterofullerenes with a metal patch using transmission electron microscopy will be implemented in the near future.

**Supporting Information Available:** Movies based on the MD simulations of the graphene-fullerene transformation for the free $C_{96}$ and $C_{384}$ graphene flakes at temperature 3000 K. Movies based on the MD simulations of the Ni-assisted graphene-fullerene transformation for the $C_{54}$ flake with the $Ni_5$ cluster attached, the $C_{96}$ flake with the $Ni_{13}$ cluster attached and the $C_{384}$ flake with the $Ni_{79}$ cluster attached at temperatures 2300 K, 2300 K and 2500 K, respectively. This material is available free of charge via the Internet at http://pubs.acs.org.

**ACKNOWLEDGEMENT** This work has been supported by the RFBR grants 11-02-00604 and 12-02-90041-Bel. The atomistic calculations are performed on the SKIF MSU Chebyshev supercomputer, MVS-100K supercomputer at the Joint Supercomputer Center of the Russian Academy of Sciences and on the Multipurpose Computing Complex "Kurchatov Institute".